\documentclass{article}
\usepackage{ijcai24}

\usepackage{soul}
\usepackage{times}
\usepackage{soul}
\usepackage{url}
\usepackage[hidelinks]{hyperref}
\usepackage[utf8]{inputenc}
\usepackage[small]{caption}
\usepackage{graphicx}
\usepackage{amsmath}
\usepackage{amsthm}
\usepackage{booktabs}
\usepackage{algorithm}
\usepackage{algorithmic}
\usepackage[switch]{lineno}
\usepackage{multirow}
\usepackage{circledsteps}
\usepackage{pifont}

\title{AVIN-Chat: An Audio-Visual Interactive Chatbot System with \\Emotional State Tuning}

\author{
Chanhyuk Park\thanks{: Equal contribution}\and
Jungbin Cho\footnotemark[1]\and
Junwan Kim\footnotemark[1]\and
Seongmin Lee \and
Jungsu Kim \\ \And
Sanghoon Lee\thanks{: Corresponding author}\\
\affiliations
Yonsei University\\
\emails
\{hanul500, whwjdqls99, jw1510, lseong721, integer528, slee\}@yonsei.ac.kr,\\
}

\begin{document}

\maketitle

\begin{abstract}
This work presents an audio-visual interactive chatbot (AVIN-Chat) system that allows users to have face-to-face conversations with 3D avatars in real-time.
Compared to the previous chatbot services, which provide text-only or speech-only communications, the proposed AVIN-Chat can offer audio-visual communications providing users with a superior experience quality.
In addition, the proposed AVIN-Chat emotionally speaks and expresses according to the user's emotional state.
Thus, it enables users to establish a strong bond with the chatbot system, increasing the user's immersion.
Through user subjective tests, it is demonstrated that the proposed system provides users with a higher sense of immersion than previous chatbot systems.
The demonstration video is available at \textcolor{magenta}{\url{https://www.youtube.com/watch?v=Z74uIV9k7_k}}.

\end{abstract}

\section{Introduction}
With the recent advancements of large language models (LLMs) \cite{brown2020language}, interactive AI chatbot services have gained attention from various industries including customer support, virtual assistants, and education. These services provide users with great convenience but their text-only{\footnote{https://juji.io/}} or audio-only{\footnote{https://assistant.google.com/}} interface hinders immersive user experience. 

\begin{figure}[h]
\centering
   \includegraphics[width=0.9\linewidth, keepaspectratio]{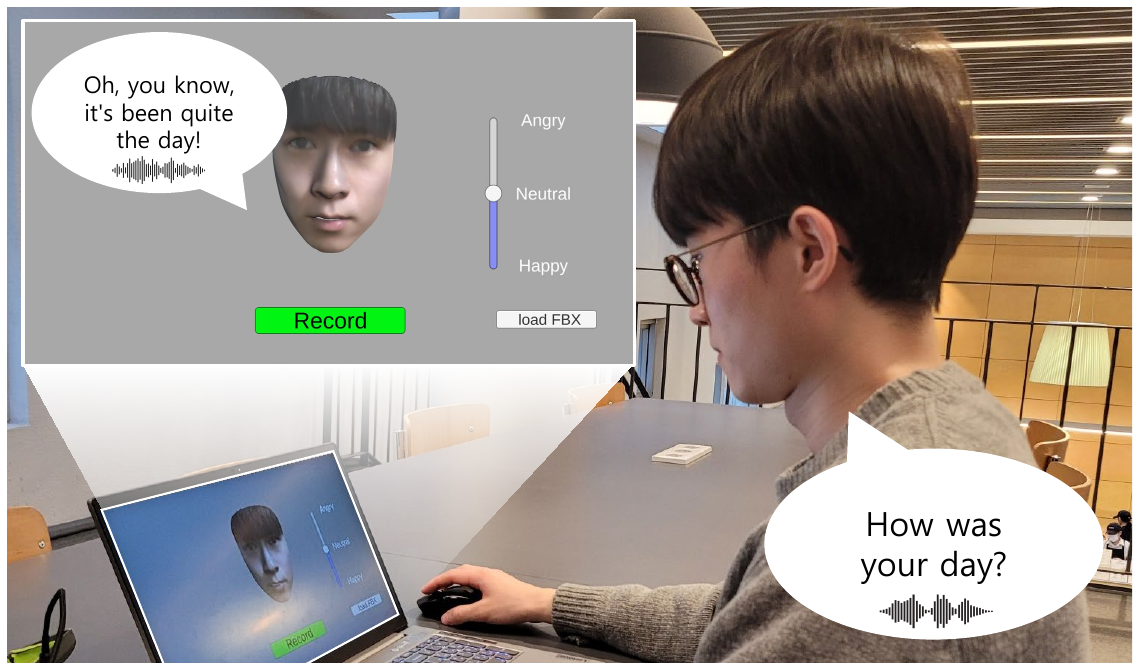} 
   \vspace{-5pt}
   \caption{Example of the AVIN-Chat use. AVIN-Chat receives the user's speech and generates an audio-visual response in real-time. 
   }
   \vspace{-5pt}
\label{fig:intro}
\end{figure}

To increase user immersion in these chatbot services, as shown in Fig. \ref{fig:intro}, we present an audio-visual interactive AI chatbot system, which listens and responds to users with a realistic and emotional talking face and voice.
To build this system, we composed our system with three distinctive sub-modules: (1) construct a deformable realistic face, (2) listen and talk to users in real-time, and (3) generate the avatar's facial expressions and lip movement in sync with its response. 
Additionally, we enable our system with emotional expressions so that it can adapt to users' feedback during conversations providing an immersive experience.

Current realistic face reconstruction methods are capable of generating realistic facial mesh along with high-quality textures \cite{Zheng2023pointavatar,Zielonka2022InstantVH,kang2021competitive,kang2020uv,lee2023faceclone} but require expensive computational costs for animation.
Thus, these methods cannot be applied to the real-time chatbot system.
Note that as we focus on animating arbitrary facial meshes, we use a template-free 3D face instead of a template-based 3D face model.
To enable template-free real-time 3D face animation, we generate facial blendshapes for expression animation from the reconstructed facial mesh.
Once blendshapes are defined, it is possible to express facial expressions with only a single linear calculation, so 3D faces can be animated with no latency \cite{lee2023stabilized,lee2023video}.


\begin{figure*}
    \centering
    \includegraphics[width=1.0\linewidth]{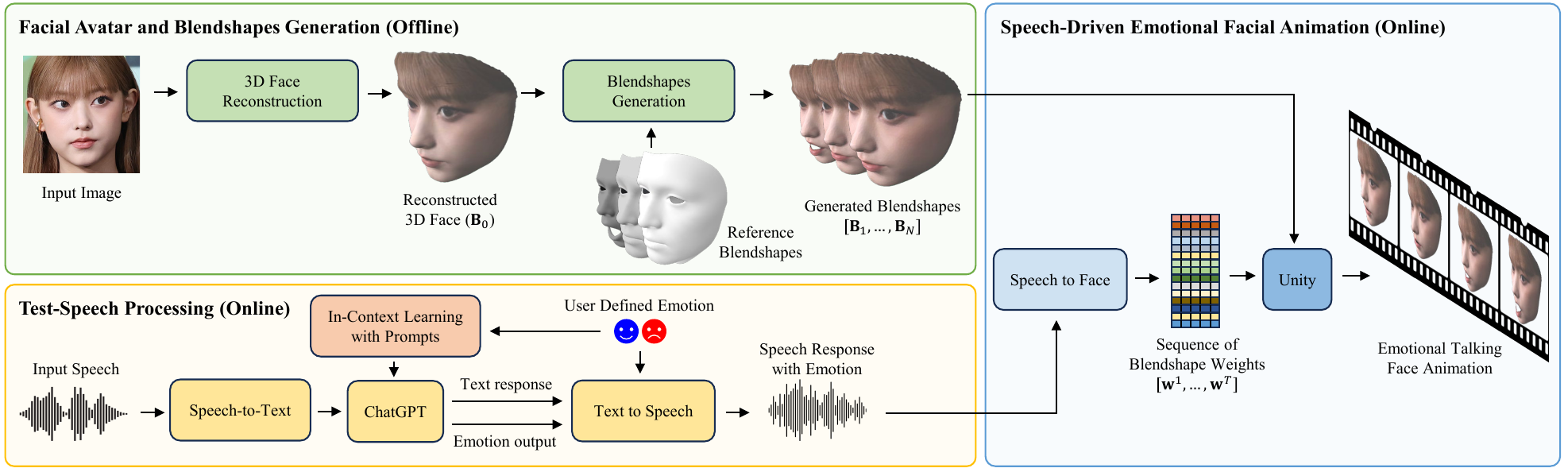}
   \vspace{-5pt}
    \caption{Overall pipeline of the proposed audio-visual interactive chatbot (AVIN-Chat) system. AVIN-Chat is constructed with three sub-modules: 1) facial avatar and blendshapes generation, 2) text-speech processing, and 3) speech-driven emotional facial animation modules.}
   \vspace{-5pt}
    \label{fig:enter-label}
\end{figure*}

For a speech-driven interactive chatbot system, we employ the speech-to-text (STT) and text-to-speech (TTS) models.
The user's speech is converted into text using the STT model and then fed to the ChatGPT{\footnote{https://openai.com/blog/chatgpt}, which is the state-of-the-art LLM.
ChatGPT acts like a brain in listening to users and generating answers in the proposed system.
Thus, after the user's statement is fed, ChatGPT generates answers and speaks using the TTS model.
Here, we perform in-context learning on ChatGPT to generate the emotional response according to the user-defined emotions.
For example, if the user wants to communicate with happy emotions, the user can adjust the ChatGPT to generate a response with happy emotions.
Finally, to animate 3D faces synchronize with the ChatGPT's response, we employ EmoTalk, which is the speech-driven 3D face animation model \cite{peng2023emotalk}.
EmoTalk generates facial expressions and lip movements from the speech input by estimating the coefficient of the facial blendshapes.
As we already built the facial blend shape that aligns with those of EmoTalk, facial expressions and lip movements can be efficiently animated with no latency.

Therefore, we present an audio-visual interactive chatbot (AVIN-Chat) system that allows users to communicate with one avatar who has the face and voice they want.
Through the subjective test, we evaluate whether the proposed audio-visual chatbot system enables intimate communication compared to other text-only or audio-only AI chatbot systems.





\section{Audio-Visual Interactive Chatbot}

\subsection{Overall Pipeline}
Figure \ref{fig:enter-label} illustrates the overall pipeline of the proposed AVIN-Chat system. 
The proposed system is composed of facial avatar generation, text-speech interaction, and speech-driven facial animation.
Subsequent sections will provide detailed descriptions of each module.

\subsubsection{Facial Avatar and Blendshapes Generation (Offline)}
The goal of facial avatar generation is to reconstruct a realistic 3D facial mesh with high-quality texture and then generate its blendshapes for efficient facial animation.
We first reconstruct a realistic facial mesh and then generate the 52 standard blendshapes allowing it to be deformable. 
For the facial reconstruction, we adopt the HRN from \cite{lei2023hierarchical}. HRN decouples facial geometry into (1) the low-frequency part for a coarse shape of the overall facial mesh, (2) the mid-frequency for local shapes, and (3) the high-frequency part for detailed information such as wrinkles. This hierarchical fashion allows us to reconstruct a photo-realistic facial mesh from a single image. 
Note that although we used a template-based 3D face reconstruction approach, we assume that the reconstructed face is a non-template face mesh for system scalability.

To make the reconstructed non-template facial mesh deformable, we generate blendshapes based on the 52 facial action units\footnote{https://arkit-face-blendshapes.com}. 
For the blendshapes generation, we employ the LDT \cite{onizuka2019landmark} that deforms the facial mesh according to the target expressions using landmarks as guidance.
Here, since LDT requires facial landmarks for facial deformation, we extract landmarks from HRN reconstructed faces.
We extract 52 facial landmarks in our implementation.
If we use the face of a character who is difficult to automatically extract landmarks, we should manually designate the landmarks.

 
\subsubsection{Text-Speech Processing (Online)}
Text-speech interaction is to listen user's speech and generate a response.
To generate the response we employ the OpenAI's ChatGPT, which is the state-of-the-art LLM.
To feed the user's speech to the ChatGPT, we first convert speech to text using the state-of-the-art STT model named Whisper \cite{radford2023robust}.
Whisper exhibits remarkable robustness in dealing with different audio conditions, effectively managing challenges such as background noise and various speech accents.
From the encoded text, ChatGPT generates the text response.
Here, to generate natural-sounding responses, we perform in-context learning on ChatGPT using prompts.

Since the ChatGPT generates text responses, we adopt the TTS model to convert text to speech.
For TTS, we utilize EmotiVoice\footnote{https://github.com/netease-youdao/EmotiVoice}, an improved version of PromptTTS \cite{guo2023prompttts}.
EmotiVoice allows us to adjust the emotion in the voice by providing certain text prompts.
The details for the ChatGPT and EmotiVoice prompts are described in Sec. \ref{sec:finetune}.




\subsubsection{Speech-Driven Emotional Facial Animation (Online)}
We employ the speech-driven facial animation model to generate the emotional talking faces from the ChatGPT's emotional speech response.
For the emotional talking face animation, we utilize the EmoTalk \cite{peng2023emotalk}.
EmoTalk estimates a sequence of blendshapes weights $[\textbf{w}^1, \dots \textbf{w}^T]$ where $\textbf{w}=[w_1^T, \dots, w_N^T]$ is the set of the $N$-blendshape and $T$ is the number of frames.
Thus, the facial shape of $t^{th}$ frame $\textbf{S}^t$ is computed as $\textbf{S}^t = \textbf{B}_0 + \textbf{w}^t (\textbf{B} - \textbf{B}_0),$
where $\textbf{B}_0$ is the reconstructed neutral face and $\textbf{B}=[\textbf{B}_1, \dots, \textbf{B}_N]$ is the generated $N$-blendshapes.
Generated face sequences $[\textbf{S}^1, \dots, \textbf{S}^T]$ are visualized using Unity.

    
        
\subsection{In-Context Learning with Prompts}
\label{sec:finetune}
It is shown that prompts significantly influence the output of LLMs \cite{schick2021exploiting}.
Therefore, there have been many attempts to control the results of LLMs through various prompting techniques \cite{shin2020autoprompt,lester2021power}. 
We utilized multiple prompts to build a chatbot system capable of natural conversation. 
In the proposed system, since ChatGPT preserves the context of the conversation, we set various conditions through prompts before starting the conversation to maintain consistency in the subsequent dialogue. To give the user a more human-like experience, we ensured the responses were not too long, avoided repetition of the same phrases, and defined do's and don'ts for responding in a friendly manner. Additionally, we informed ChatGPT that the response text would be converted to speech, instructing it not to generate responses that are hard to verbalize, like emojis.

Furthermore, we design our system so that users can input their desired emotional state.
Using the user interface in Unity, the user can easily select the desired emotional state and the selected emotion is fed to the ChatGPT and EmotiVoice as prompts.
If users are not satisfied with the chatbot's current emotional state, they can simply adjust the chatbot's emotional state during the conversation using the graphical user interface (GUI) in Unity.
By implementing this approach, we observed that the generated responses were significantly more cheerful and friend-like compared to previous answers.
It implies that the proposed chatbot is actively engaging in leading the conversation.

\subsection{Implementation Details}
\label{sec:implementation}
The overall structure of the AVIN-Chat consists of a Python backend server and a Unity frontend. In the facial avatar generation module, Unity sends an image captured by a webcam to the server. Then using the HRN, the server generates a 3D facial mesh in OBJ format along with a texture file. Using the OBJ file as a base, the LDT model creates 52 additional OBJ files, and the Blender Python API combines them to create an FBX file, which is the final blendshapes. Unity receives this 3D FBX with its texture file and displays the constructed face to the user.
In the conversational part, Unity records the user's voice, sending it to the server. In return, Unity receives an answering voice along with a list of blendshapes parameters. Once this exchange is done, Unity runs the audio and animates the facial mesh with the blendshapes parameters. For the emotional tuning, we've configured Unity to communicate with a backend API to change ChatGPT's prompts to match the user-defined emotions. Details of our system's graphical user interface are shown in Fig. \ref{fig:GUI}.
We used AMD Ryzen 7 5800u and AMD Radeon (TM) Graphics for the client and AMD Ryzen 5 5600X and NVIDIA RTX 3060 Ti for the server.

\begin{figure}[h]
\centering
   \includegraphics[width=1.0\linewidth]{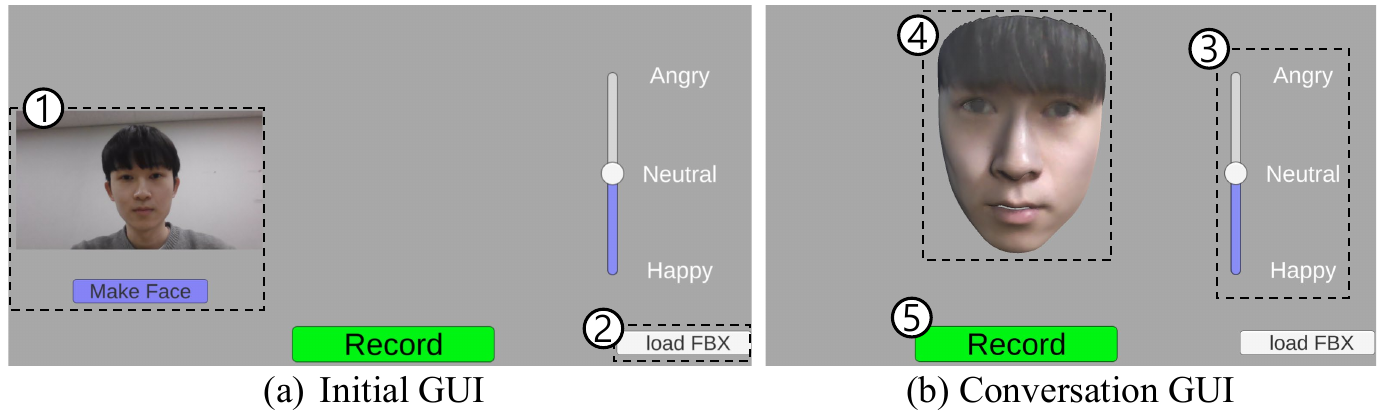} 
   \vspace{-15pt}
   \caption{GUI visualization and example of actual use of the AVIN-Chat: (a) Initial GUI where users can \ding{172} capture an image or \ding{173} load an FBX file, and (b) conversation GUI where users can \ding{174} define emotions and talk to \ding{175} facial avatar by using the \ding{176} record button.
   }
   \vspace{-10pt}
\label{fig:GUI}
\end{figure}


\section{Experimental Results}
We evaluate our proposed system on what conversations users prefer with different AI chatbots, such as text-only and speech-only systems, compared to ours. The text-only system utilizes ChatGPT which interacts with the user only through text inputs and outputs. 
The speech-only system is composed of ChatGPT, Whisper, and EmotiVoice.
Note that the speech-only system also can generate emotional answers because it utilizes EmotiVoice.
For fair comparisons, we use the same model weights as those used in the proposed AVIN-Chat. 
We recruited 11 participants to score each system's preference over the proposed AVIN-Chat.
When scoring, participants are asked to measure their preferences in terms of intimacy, immersiveness, empathy, and overall satisfaction.
Figure \ref{fig:user_preference} shows the subjective test results.
It shows that the text-only chatbot achieves better performance than the speech-only chatbot.
Participants reported that visual stimulus, although being only text, enhanced user experience more than voice stimulus.
In addition, the results demonstrate that the proposed AVIN-Chat significantly outperforms other chatbot systems in terms of preference by providing both visual and auditory stimulus.

\begin{figure}[h]
\centering
   \includegraphics[width=1.0\linewidth]{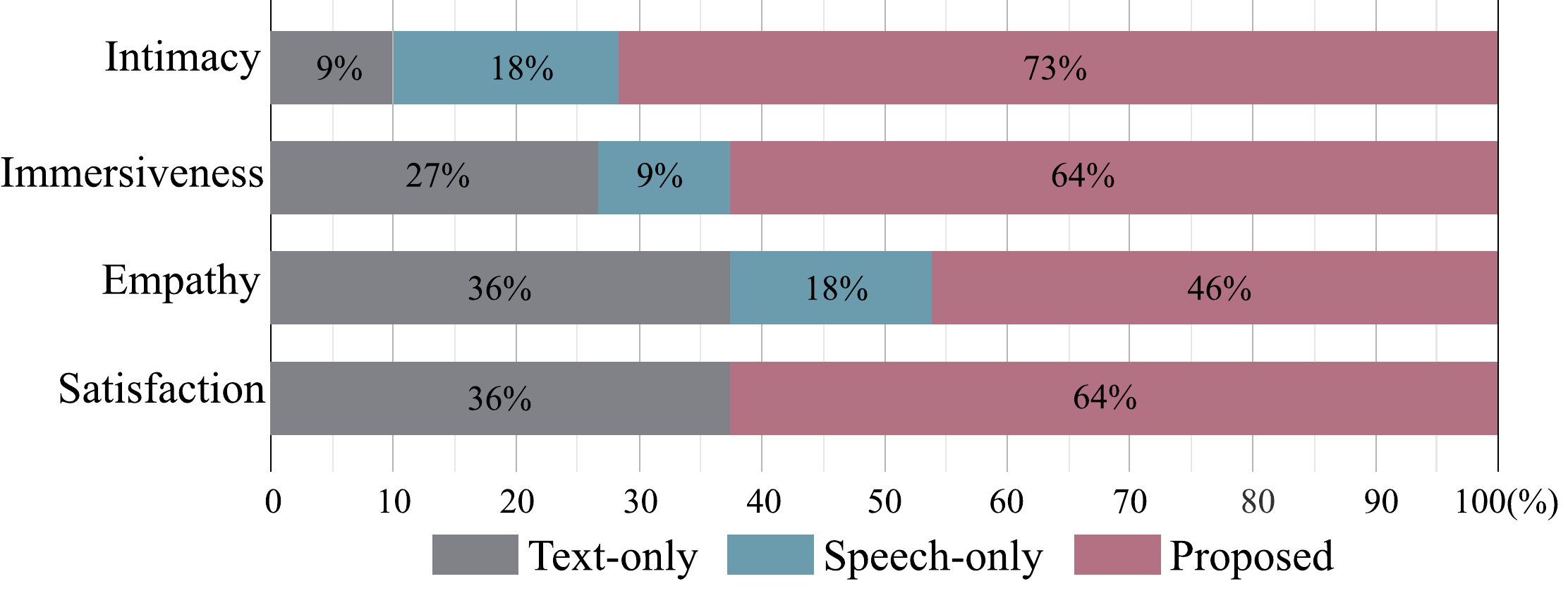} 
   \vspace{-10pt}
   \caption{User preference scores for different chatbot systems on intimacy, immersiveness, empathy, and overall satisfaction.}
   \vspace{-10pt}
\label{fig:user_preference}
\end{figure}

\section{Conclusion}
In this paper, we have presented an audio-visual interactive chatbot, an end-to-end system for having face-to-face conversations with facial avatars reconstructed from a single image.
Experimental results demonstrate that the proposed system can provide more immersive communication compared to text-only and speech-only chatbot systems.
We believe that the proposed system can encourage users to use chatbot services and can be applied to various applications such as education and medical therapy.
Our future work is to incorporate body gestures into a talking face for more realistic virtual communications.

\section{Acknowledgements}
This research was supported by Culture, Sports and Tourism R\&D Program through the Korea Creative Content Agency grant funded by the Ministry of Culture, Sports and Tourism in 2024 (Project Name: Global Talent for Generative AI Copyright Infringement and Copyright Theft, Project Number: RS-2024-00398413, Contribution Rate: 100\%).


\bibliographystyle{named}
\bibliography{ijcai24}

\begin{thebibliography}{}

\bibitem[\protect\citeauthoryear{Brown \bgroup \em et al.\egroup }{2020}]{brown2020language}
Tom Brown, Benjamin Mann, Nick Ryder, Melanie Subbiah, Jared~D Kaplan, Prafulla Dhariwal, Arvind Neelakantan, Pranav Shyam, Girish Sastry, Amanda Askell, et~al.
\newblock Language models are few-shot learners.
\newblock {\em Advances in neural information processing systems}, 33:1877--1901, 2020.

\bibitem[\protect\citeauthoryear{Guo \bgroup \em et al.\egroup }{2023}]{guo2023prompttts}
Zhifang Guo, Yichong Leng, Yihan Wu, Sheng Zhao, and Xu~Tan.
\newblock Prompttts: Controllable text-to-speech with text descriptions.
\newblock In {\em ICASSP 2023-2023 IEEE International Conference on Acoustics, Speech and Signal Processing (ICASSP)}, pages 1--5. IEEE, 2023.

\bibitem[\protect\citeauthoryear{Kang \bgroup \em et al.\egroup }{2020}]{kang2020uv}
Jiwoo Kang, Seongmin Lee, and Sanghoon Lee.
\newblock Uv completion with self-referenced discrimination.
\newblock In {\em Eurographics (Short Papers)}, pages 61--64, 2020.

\bibitem[\protect\citeauthoryear{Kang \bgroup \em et al.\egroup }{2021}]{kang2021competitive}
Jiwoo Kang, Seongmin Lee, and Sanghoon Lee.
\newblock Competitive learning of facial fitting and synthesis using uv energy.
\newblock {\em IEEE Transactions on Systems, Man, and Cybernetics: Systems}, 52(5):2858--2873, 2021.

\bibitem[\protect\citeauthoryear{Lee \bgroup \em et al.\egroup }{2023a}]{lee2023faceclone}
Kyungjune Lee, Jeonghaeng Lee, Hyucksang Lee, Mingyu Jang, Seongmin Lee, and Sanghoon Lee.
\newblock Faceclone: Interactive facial shape and motion cloning system using multi-view images.
\newblock In {\em 2023 IEEE International Conference on Multimedia and Expo Workshops (ICMEW)}, pages 512--513. IEEE, 2023.

\bibitem[\protect\citeauthoryear{Lee \bgroup \em et al.\egroup }{2023b}]{lee2023video}
Seongmin Lee, Hyunse Yoon, Jiwoo Kang, Jungsu Kim, Jiwan Son, Jungwoo Huh, and Sanghoon Lee.
\newblock Video-based stabilized 3d face alignment using temporal multi-discrimination.
\newblock In {\em 2023 IEEE 25th International Workshop on Multimedia Signal Processing (MMSP)}, pages 1--6. IEEE, 2023.

\bibitem[\protect\citeauthoryear{Lee \bgroup \em et al.\egroup }{2023c}]{lee2023stabilized}
Seongmin Lee, Hyunse Yoon, Sohyun Park, Sanghoon Lee, and Jiwoo Kang.
\newblock Stabilized temporal 3d face alignment using landmark displacement learning.
\newblock {\em Electronics}, 12(17):3735, 2023.

\bibitem[\protect\citeauthoryear{Lei \bgroup \em et al.\egroup }{2023}]{lei2023hierarchical}
Biwen Lei, Jianqiang Ren, Mengyang Feng, Miaomiao Cui, and Xuansong Xie.
\newblock A hierarchical representation network for accurate and detailed face reconstruction from in-the-wild images.
\newblock In {\em Proceedings of the IEEE/CVF Conference on Computer Vision and Pattern Recognition}, pages 394--403, 2023.

\bibitem[\protect\citeauthoryear{Lester \bgroup \em et al.\egroup }{2021}]{lester2021power}
Brian Lester, Rami Al-Rfou, and Noah Constant.
\newblock The power of scale for parameter-efficient prompt tuning.
\newblock In {\em Proceedings of the 2021 Conference on Empirical Methods in Natural Language Processing}, pages 3045--3059, 2021.

\bibitem[\protect\citeauthoryear{Onizuka \bgroup \em et al.\egroup }{2019}]{onizuka2019landmark}
Hayato Onizuka, Diego Thomas, Hideaki Uchiyama, and Rin-ichiro Taniguchi.
\newblock Landmark-guided deformation transfer of template facial expressions for automatic generation of avatar blendshapes.
\newblock In {\em Proceedings of the IEEE/CVF International Conference on Computer Vision Workshops}, pages 0--0, 2019.

\bibitem[\protect\citeauthoryear{Peng \bgroup \em et al.\egroup }{2023}]{peng2023emotalk}
Ziqiao Peng, Haoyu Wu, Zhenbo Song, Hao Xu, Xiangyu Zhu, Jun He, Hongyan Liu, and Zhaoxin Fan.
\newblock Emotalk: Speech-driven emotional disentanglement for 3d face animation.
\newblock In {\em Proceedings of the IEEE/CVF International Conference on Computer Vision}, pages 20687--20697, 2023.

\bibitem[\protect\citeauthoryear{Radford \bgroup \em et al.\egroup }{2023}]{radford2023robust}
Alec Radford, Jong~Wook Kim, Tao Xu, Greg Brockman, Christine McLeavey, and Ilya Sutskever.
\newblock Robust speech recognition via large-scale weak supervision.
\newblock In {\em International Conference on Machine Learning}, pages 28492--28518. PMLR, 2023.

\bibitem[\protect\citeauthoryear{Schick and Sch{\"u}tze}{2021}]{schick2021exploiting}
Timo Schick and Hinrich Sch{\"u}tze.
\newblock Exploiting cloze-questions for few-shot text classification and natural language inference.
\newblock In {\em Proceedings of the 16th Conference of the European Chapter of the Association for Computational Linguistics: Main Volume}, pages 255--269, 2021.

\bibitem[\protect\citeauthoryear{Shin \bgroup \em et al.\egroup }{2020}]{shin2020autoprompt}
Taylor Shin, Yasaman Razeghi, Robert~L Logan~IV, Eric Wallace, and Sameer Singh.
\newblock Autoprompt: {E}liciting knowledge from language models with automatically generated prompts.
\newblock In {\em Proceedings of the 2020 Conference on Empirical Methods in Natural Language Processing (EMNLP)}, pages 4222--4235, 2020.

\bibitem[\protect\citeauthoryear{Zheng \bgroup \em et al.\egroup }{2023}]{Zheng2023pointavatar}
Yufeng Zheng, Wang Yifan, Gordon Wetzstein, Michael~J. Black, and Otmar Hilliges.
\newblock Pointavatar: Deformable point-based head avatars from videos.
\newblock In {\em Proceedings of the IEEE/CVF Conference on Computer Vision and Pattern Recognition (CVPR)}, 2023.

\bibitem[\protect\citeauthoryear{Zielonka \bgroup \em et al.\egroup }{2022}]{Zielonka2022InstantVH}
Wojciech Zielonka, Timo Bolkart, and Justus Thies.
\newblock Instant volumetric head avatars.
\newblock {\em 2023 IEEE/CVF Conference on Computer Vision and Pattern Recognition (CVPR)}, pages 4574--4584, 2022.

\end{thebibliography}

\end{document}